\documentclass[11pt]{article}

\usepackage[a4paper,margin=1in]{geometry}

\usepackage[T1]{fontenc}
\usepackage[scaled=0.92]{helvet}

\usepackage{setspace}
\usepackage{graphicx}
\usepackage{float}        
\usepackage{placeins}     
\usepackage{xcolor}

\usepackage{amsmath,amssymb}

\usepackage[numbers,sort&compress]{natbib}

\usepackage[hidelinks]{hyperref}

\usepackage{caption}
\usepackage[hidelinks]{hyperref}

\makeatletter
\renewcommand{\maketitle}{
  \begin{flushleft}
{\bfseries\fontsize{22}{22}\selectfont \@title \par}
    \vspace{1em}
    {\large \@author \par}
    \vspace{1em}
  \end{flushleft}
}
\makeatother

\makeatletter
\renewenvironment{abstract}{
  \begin{flushleft}
  \bfseries Abstract\par
  \vspace{0.4em}
  \normalfont
  \small
  \setstretch{1.05}
}{
  \end{flushleft}
}
\makeatother
\vspace{0.3em}

\newcommand{\keywords}[1]{
  \begin{flushleft}
  \vspace{0.5em}
  \textbf{Keywords:} #1
  \end{flushleft}
}

\title{Is segregation encoded in urban form?\\An entropy-based analysis}
\author{Vinicius M. Netto, Caio Cacholas, Camila Carvalho, Edgardo Brigatti}
\date{February 2026}

\begin{document}
\maketitle

\begin{abstract}
The footprints of residential segregation have long been documented, yet the role of urban form as both medium and manifestation of segregation remains under-specified. We investigate whether the configuration of the built fabric may encode residential segregation in its spatial structure, hypothesising that built-form entropy (BFE) regimes are associated with the spatial distribution of income groups and their local clustering in non-linear ways. We examine this by quantifying BFE through a Shannon-based measure computed from building footprints, characterising income-based distributions using the Gini index and Moran's I, and placing both on a common spatial footing through a regular tessellation. Applying this framework to S\~ao Paulo, Latin America’s largest city, we find non-linear relationships between BFE, income, and segregation: income levels and residential clustering increase toward both extremes of the entropy spectrum, with a stronger rise at the high-entropy end. This asymmetry suggests that high-entropy urban forms are associated with distinct spatial processes of segregation, including elite enclaving and incremental development in lower-income settlements, while low-entropy forms reflect more selective occupation shaped by planning and market filtering. Overall, the findings suggest that built form is more than a neutral backdrop, functioning as both affordance and signal of segregation.
\end{abstract}


\section*{Main}
The urban imprint of residential segregation has been documented in multiple ways, from early 
mappings of class- and occupation-based divisions in the Chicago School tradition \cite{burgess_residential_1928} 
to formal indices capturing unevenness, proximity, and distance-decay in group distributions (e.g. \cite{wong_formulating_2005}). Broadly defined as the spatial clustering of socially homogeneous groups, residential 
segregation has recently been identified as both the most extensively studied form of segregation 
and the most central node within the broader topology of segregation forms \cite{netto2024decoding}. The analysis of group distributions in cities dates back at least to the late nineteenth century, notably with Charles Booth’s poverty maps of London (1889) \cite{booth1889poverty} and Emily Balch’s (1895) synthesis of Hull House 
mappings in Chicago \cite{balch1895hull}. Since Duncan and Duncan’s (1955) foundational work \cite{duncan_methodological_1955}, segregation has most 
often been quantified using dissimilarity-type indices, including the Gini index as a measure of 
evenness \cite{james1985measures} and entropy-based measures \cite{theil1971note}. 
These classic indices are aspatial, relying just on population counts 
rather than on the spatial organisation of the environment itself.
Subsequent advances reframed residential segregation as a scale-dependent spatial process \cite{reardon_measures_2004}, prompting the development of measures that incorporate contiguity and distance – such as boundary- and weight-modified dissimilarity indices that account for shared borders and spatial proximity \cite{morrill1991measure, wong_spatial_1993} (see Supplementary Information for details). This trajectory reflects a broader methodological shift from descriptive mapping to aspatial quantification and, more recently, to explicitly spatial approaches. Within this progression, Moran's $I$ emerged as a prototypical measure of spatial autocorrelation, providing a standard way to assess whether similar social groups cluster in neighbouring locations beyond what would be expected under spatial randomness. In this sense, measures such as the Gini index capture the composition of income distributions within areas, whereas Moran's $I$ captures their spatial arrangement, distinguishing unevenness from the extent to which it is expressed as local clustering.\\

Yet across this progression, the physical configuration of the built environment remains largely under-specified, as spatial measures continue to capture where groups are located rather than how the morphology of urban space structures their separation and interaction. 
Even when spatial measures incorporate contiguity and distance,
they still offer limited 
insight into actual residential distributions and how groups interface along the permeable public realm – streets, squares, and other open spaces between buildings where everyday interactions 
unfold. Adjacency alone contains little information about the internal spatial structure of urban 
areas. In parallel, growing interest has turned to the potential roles of urban form in shaping 
segregation patterns. Here, it is useful to distinguish two complementary components: the street 
network (SN), the connective fabric of open spaces that structures movement and accessibility, and 
built form (BF), understood as the shape, placement, and arrangement of buildings. Although the 
street network has received considerable attention in residential segregation research \cite{vaughan2005relationship, vaughan2007spatial, saboya2025chasm}, the role of built form has remained comparatively 
under-specified, often subsumed under population distributions or generic notions of proximity. A 
key challenge, therefore, is to move beyond abstract measures of contiguity or distance and explicitly integrate the texture of the built environment – configurations of building footprints – into the 
analysis of residential segregation. \\
 
Built form may matter for residential segregation insofar as its spatial characteristics relate to where 
social groups locate and how they interact. We examine these characteristics with the lens of order and entropy in the built fabric. Built form is represented here two-dimensionally through building 
footprints, in the spirit of Nolli maps, and analysed as cellular arrangements. High built-form entropy 
denotes greater variability in these arrangements, typically observed where buildings are positioned 
with little apparent spatial order. Conversely, more ordered patterns concentrate on a narrower set 
of configurations, allowing them to find higher frequencies, increasing predictability and reducing 
entropy. Recent work suggests that entropy can encode social information associated with distinct 
spatial cultures \cite{brigatti_entropy_2021, netto_urban_2023}. 
This motivates our central research question: do configurations of built form also “encode” residential segregation?  \\
 
We hypothesise that variation in built-form entropy is associated with morphological conditions 
under which income groups cluster, as opposed to a null expectation in which the spatial distribution of built-form entropy bears no systematic relationship to income-based residential patterns. Built-form entropy 
differentiates urban areas in terms of spatial order and configuration, and may be implicated in how boundaries between groups are perceived and maintained. 
Entropy regimes may have consequences for locational residential decisions: 
intricate and irregular configurations tend to reduce intervisibility and through-movement, limiting 
access and exposure between groups, whereas more ordered configurations tend to increase legibility, predictability, and potential co-presence (cf. \cite{hillier1993natural,roberto2021spatial}). 
In this sense, built-form entropy may be associated with the spatial distribution of income groups across entropy regimes, as it relates to varying conditions of accessibility, visibility, and interaction that accompany patterns of spatial separation. 
Higher-income households may tend to locate in highly entropic areas as soft barriers to access, as intricate spatial configurations are associated with less legible and more circuitous paths for outsiders, reduced intervisibility, and constrained through-movement, while their visual complexity may also signal distinction from surrounding areas.
Conversely, lower-income communities with strong social bonds may find intricate spaces supportive of local cohesion at the expense of contact with surrounding communities \cite{hillier1984social}. 
Another route to high entropy can arise where formal planning is weak or absent, e.g. in settings of informal urbanisation, conditions more commonly associated with lower-income groups in Brazil. 
In short, income groups may – consciously or not – seek, shape, or be constrained into areas characterised by specific entropy regimes. Variation in spatial order and entropy can be mobilised in locational choices, as a morphological condition through which residential segregation emerges and persists. These considerations motivate the following complementary hypotheses: 

\begin{itemize}
    \item[$\bullet$] H1: Higher built-form entropy is associated with stronger local clustering in areas where different income groups coexist. 
        
    \item[$\bullet$] H2: The association between built-form entropy and residential segregation varies across income levels, with some groups concentrating in specific entropy regimes, particularly high-income groups in high-entropy areas.
    
    \item[$\bullet$] H3: The association of built-form entropy with residential segregation is non-linear, with segregation varying across entropy values.
 \end{itemize}
 
Taken together, these hypotheses operationalise the proposed framework by articulating 
complementary dimensions of how built-form entropy relates to residential segregation. 
H1 specifies the overall direction and intensity of the association, linking higher entropy to stronger local clustering. In other words, where the physical structure of an area is more heterogeneous, people with similar incomes may be more likely to cluster locally than mix evenly. Neighbourhoods with more diverse and entropic built forms likewise tend to show stronger clustering of income groups.
H2 recognises that income groups are unevenly distributed along the built-form entropy spectrum. 
In practice, high-income groups tend to concentrate in high-entropy areas, middle-income groups are more broadly distributed across the spectrum, and lower-income groups are also more frequently found in relatively entropic contexts. 
This may also result in socially homogeneous areas with low internal inequality and clustering.
H3 further refines this framework by positing a non-linear relationship, in which segregation varies across the entropy spectrum rather than following a monotonic trend. Segregation does not simply increase or decrease as urban form becomes more entropic; instead, it follows different patterns across the entropy spectrum, with intermediate-entropy contexts differing systematically from both lower- and higher-entropy areas. 
H1 addresses high-entropy contexts in which multiple groups coexist and cluster locally, H2 addresses the concentration of income groups in certain entropy regimes, and H3 addresses the non-linearity of the relationship between built-form entropy and residential segregation.
\\

\begin{figure}[t] 
    \centering 
    \includegraphics [width=\linewidth]{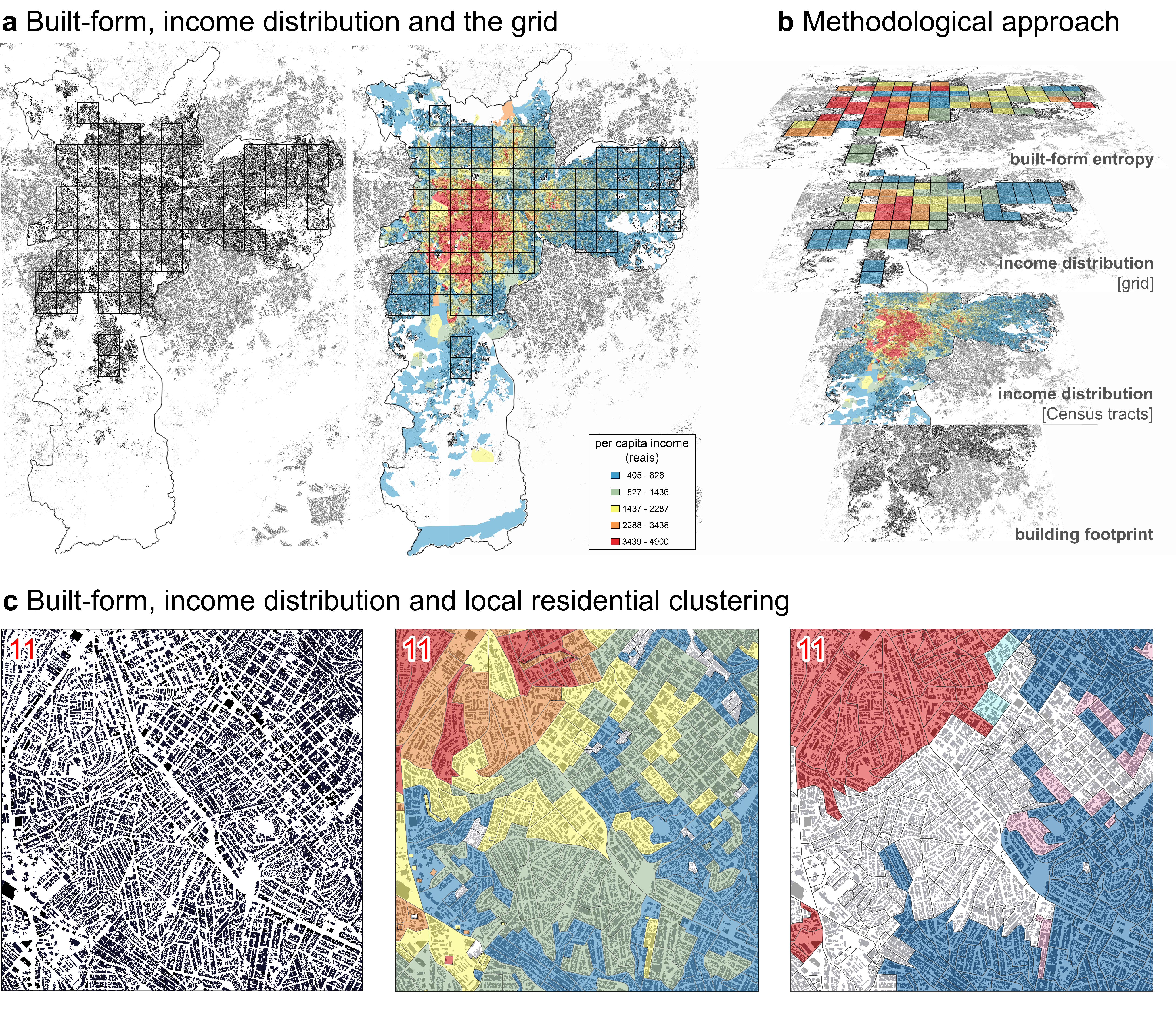} 
    \caption{\small (a) Building footprints, tract-level average per-capita income and the regular 3×3 km 
    grid. (b) Method schematic: overlapping the footprint and income layers to compute built-form entropy and residential-segregation metrics on the grid. (c) Grid-cell detail showing high built-form entropy, tract-level per-capita income, and Moran's $I$ (spatial autocorrelation). 
    Sources: OpenStreetMap/GeoSampa (building footprint), IBGE (census tracts data), authors (analysis).} 
    \label{fig:Fig_1} 
\end{figure}

We aim to contribute by adding built form as a missing spatial component in residential segregation research. Rather than directly testing causal mechanisms, the analysis examines systematic spatial alignments between 
built-form configurations and income-based residential patterns, raising questions about how urban 
morphology may be implicated in segregation dynamics (Fig.~\ref{fig:Fig_1}).
We proceed in three steps. 
First, we quantify randomness in the built fabric using a corrected 
Shannon entropy computed from building-footprint configurations. 
Second, we characterise income-based residential 
patterns using census-tract data by observing citywide gradients in average per-capita income, and, at the local scale within grid cells, analysing inequality with Gini index and quantifying clustering using Moran's $I$. 
While a segregation metric is applied on the local (cell) scale, the broader income distribution of S\~ao Paulo provides an essential contextual backdrop. 
Third, we examine how built-form entropy relates to these local 
measures across the urban territory. 
We apply this framework to the S\~ao Paulo metropolitan area, the largest in Latin America.

\section*{Results}

\begin{figure}[t] 
    \centering 
    \includegraphics [width=\linewidth]{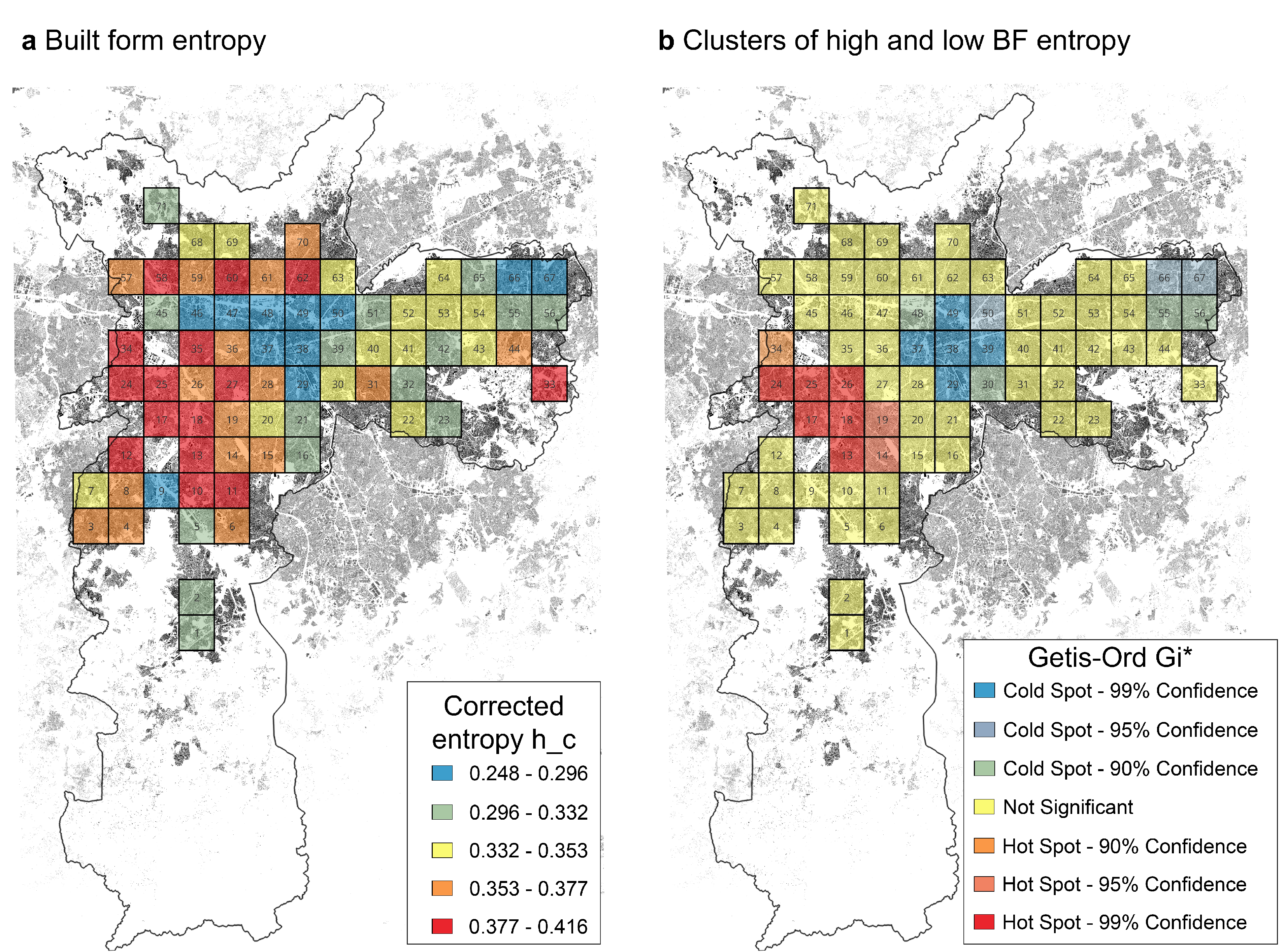} 
    \caption{\small (a) Building footprint (background) and entropy estimation. The urban fabric exhibits spatial continuity in estimated $h_c$. (b) Hot Spot Analysis of BFE 
    with values computed for the 3×3 km grid cells. Red cells indicate statistically significant clusters of high entropy (hot spots), and blue cells indicate clusters of low entropy (cold spots).
} 
    \label{fig:Fig_2} 
\end{figure}

Figure \ref{fig:Fig_2} shows 
corrected-entropy estimates for the square grid of S\~ao Paulo. We observe spatial continuity in the values of $h_c$, delineating areas with shared morphological characteristics. Corrected entropy exhibits clusters of similar values, with low entropy across the central area (cells 37 and 38) – including parts of districts Rep\'ublica, S\'e, Bela Vista, Bras, Mooca and Cambuci – and high entropy in districts further south, surrounding cell 13 (e.g. Jaguaré and Santo Amaro – see the complete values per cell for corrected entropy, income, Gini and Moran's $I$ in the Supplementary Information).

\begin{figure}[!htbp] 
    \centering 
    \includegraphics [width=\linewidth]{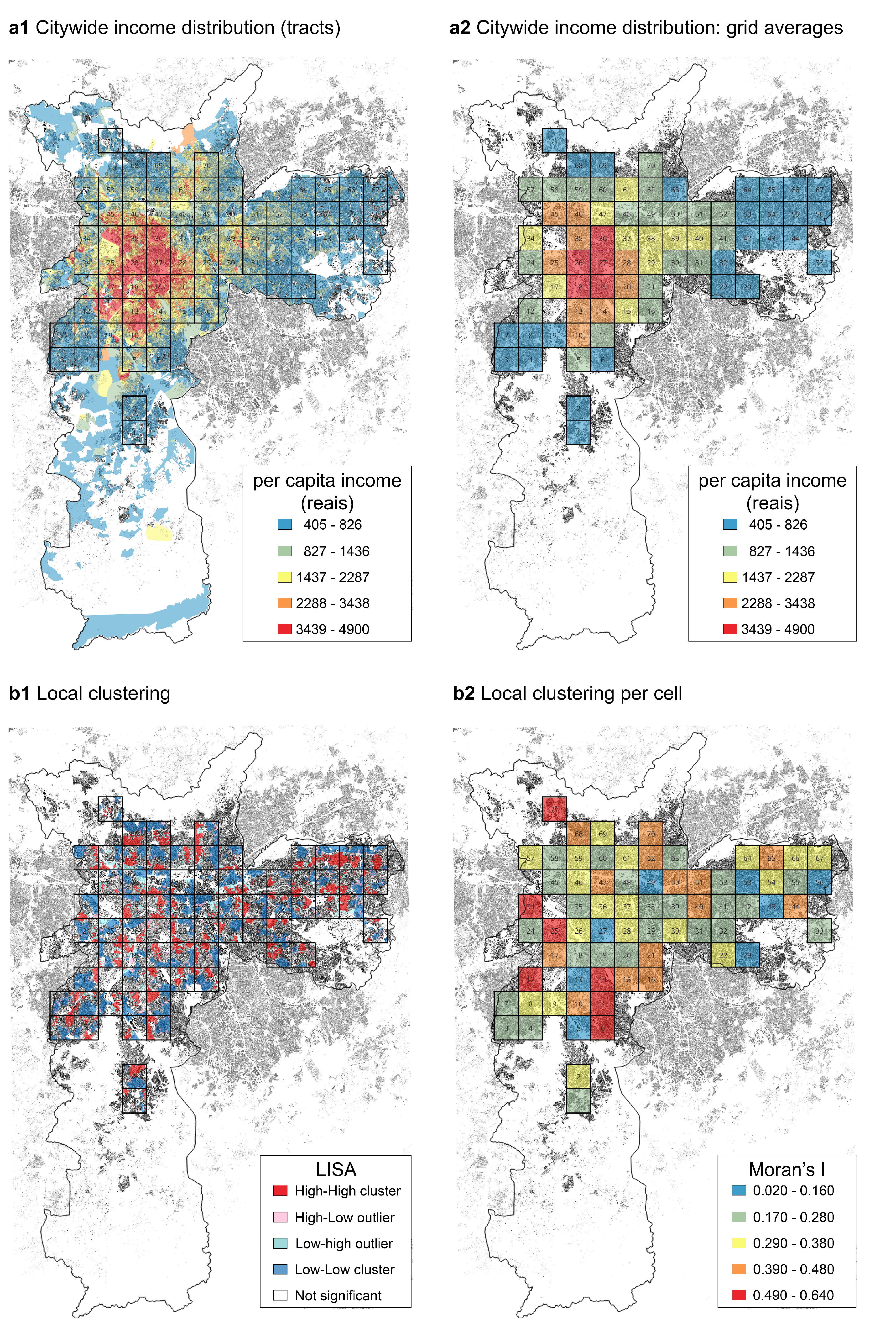} 
    \caption{\small Income and residential segregation in S\~ao Paulo: (a1) The citywide distribution of per-capita income by census tract shows a remarkable centre–periphery pattern; (a2) per capita income distribution averaged at the grid cell scale; (b1) Moran's $I$ computed per each 3×3 km grid cell; (b2) results for Moran's $I$.} 
    \label{fig:Fig_3} 
\end{figure}

Figure \ref{fig:Fig_3} shows the values of per capita income, Moran's $I$, and the Gini index for each square. Per capita income provides explicit locational information on social groups, derived by 
areal-weighting census-tract values onto the square grid. It exhibits a clear centre–periphery pattern: 
high values cluster in central districts and decline toward the urban fringe, with a broad, relatively uniform low-income area in the east.

In turn, Gini provides the degree of income inequality between the average per capita income for census tracts within each grid cell. Values are low in the east, where high- and middle-income groups are largely absent, and higher in the area south to (and including) the city centre, where income levels are more mixed. Finally, Moran's $I$ provides an estimation of the spatial autocorrelation using census-tracts values areal-weighted onto the grid cells. In S\~ao Paulo, Moran's $I$ displays considerable variation across space and, therefore, requires a more nuanced interpretation. 

\subsection*{Non-linear relationships between built-form entropy and segregation} 
\begin{figure}[!htbp] 
    \centering 
    \includegraphics [width=\linewidth]{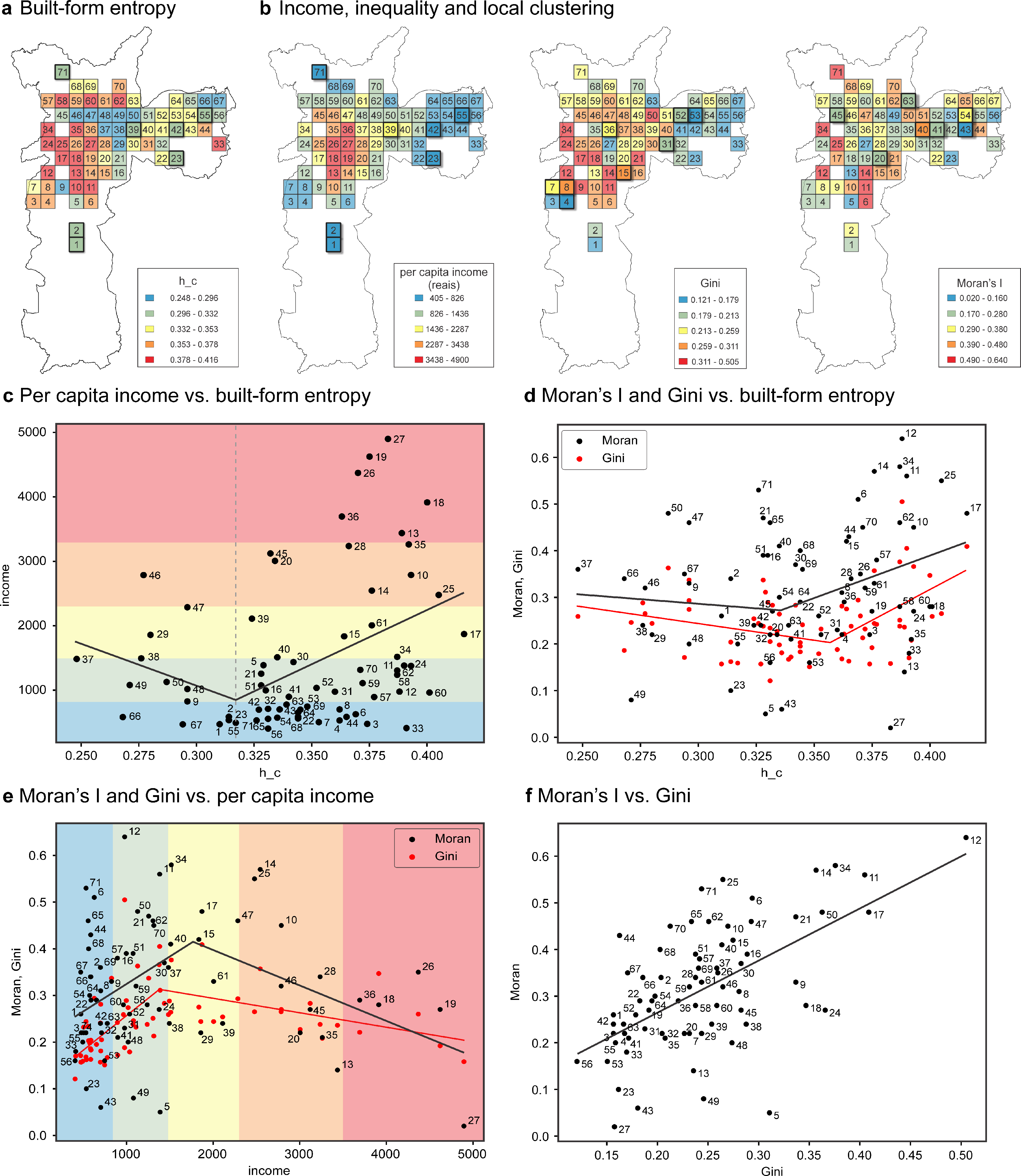} 
    \caption{\small Built-form entropy, income distribution and local residential segregation in S\~ao Paulo. Top panels: (a) corrected built-form entropy ($h_c$) per grid cell; (b) average per-capita income, income inequality (Gini) and residential segregation (Moran's $I$) on the same grid. Cells outlined with thicker black lines in (a) and (b) mark areas whose values lie near the estimated threshold. (c) Scatterplot of per-capita income versus $h_c$: the solid line shows the piecewise regression; the vertical dashed line identifies the estimated entropy threshold $h^T_c\sim 0.317$. (d) $h_c$ versus Moran's $I$ (black) and Gini (red); solid line, piecewise regression. Cell numbers corresponding to Gini values are reported in Fig. S2 in the Supplementary Information. (e) Income versus Moran's $I$ (black) and Gini (red); solid lines, piecewise regression. (f) Moran's $I$ versus Gini; solid line, best linear fit.} 
    \label{fig:Fig_4} 
\end{figure}

Having estimated built-form entropy ($h_c$) and residential segregation, we ask how they relate. If 
segregation is intertwined with urban morphology, our entropy measure should capture part of that 
relationship by quantifying the randomness in building-footprint configurations. We therefore examine whether systematic spatial associations emerge between built-form entropy and residential segregation across S\~ao Paulo. 
At first glance, these variables appear to be only weakly related. Yet, the relationship clearly displays a threshold behaviour.
We begin by examining the scatterplot of the corrected BF entropy against income (Fig. \ref{fig:Fig_4}a-b). 
 
For entropy values below a threshold, the corresponding income is generally moderate or low; by contrast, above that threshold, very high incomes become more likely. We quantify this pattern with 
a piecewise (segmented) regression \cite{toms2003piecewise}. Specifically, we fit a simple broken-stick model – two linear segments joined at an unknown breakpoint – where the estimated breakpoint defines the $h_c$ threshold. The best-fitting broken-stick model yields a first segment 
with a negative slope and a second with a positive one. At low $h_c$, income is moderate and 
declines as $h_c$ increases. Once the estimated threshold $h^T_c\sim 0.317$ is reached, the 
relationship reverses: beyond this threshold,  higher $h_c$ produce an increase in income values. In short, the association is negative below the threshold and positive above it, with income attaining its minimum around $h^T_c$.\\

An analogous threshold behaviour can be observed when plotting the Gini or Moran's $I$ as a function of entropy (Fig.~\ref{fig:Fig_4}e). The estimated breakpoint again lies near the median $h_c$. Below the 
threshold, entropy is likely to cause a slight decrease in both indices, which reach their lowest values at the threshold. Above it, both indices increase. The parallel behaviour is consistent with the strong correlation between Gini and Moran's $I$ (see Fig.~\ref{fig:Fig_4}f). The Gini index clearly increases with Moran's $I$, indicating that higher income inequality is accompanied by stronger spatial segregation. 
Finally, we analysed how Gini and Moran's $I$ vary with income. Both exhibit a threshold behaviour: 
the fitted piecewise regressions have a positive slope at lower incomes and a negative slope beyond 
the  estimated breakpoint. Most of the highest values of both indices occur below the breakpoint, while at higher incomes they decline (Fig.~\ref{fig:Fig_4}e). In other words, above the income threshold, higher 
incomes coincide with the absence of high Gini and Moran's $I$ values, indicating greater income equality across the tracts composing each grid cell and weaker local clustering relative to neighbouring cells. Together, these patterns indicate a systematic, non-linear spatial alignment between built-form entropy, income, and residential clustering.

\section*{Discussion}

In line with accounts of residential segregation as a scale-dependent phenomenon \cite{johnston2016macro, fowler2016segregation, catney2018complex} and with location theory since Alonso \cite{alonso1964location}
 (e.g. \cite{brueckner1999central, guerrieri2013endogenous}), S\~ao Paulo shows a pronounced centre–periphery pattern of income-based residential distribution shaped by citywide locational forces, alongside more localised manifestations of segregation expressed as clustering of different income groups within grid cells. 
These two expressions are analytically distinct: large-scale income concentrations can dominate entire areas without necessarily generating high levels of within-cell spatial clustering, whereas local clustering may emerge even within broader income gradients. 
Against this backdrop, we hypothesised that variation in built-form entropy is associated with the morphological conditions under which income groups cluster locally. 
Specifically, in areas where different income groups coexist, higher built-form entropy tends to be associated with stronger local clustering (H1); income groups are unevenly distributed across the entropy spectrum, with some groups concentrating in specific entropy regimes
(H2);
the associations of built-form entropy, income and residential segregation are non-linear 
(H3). 
To clarify how the hypotheses map onto the empirical patterns, we identify a set of recurrent regimes
in the wider continuum of cases associated with the threshold-based relationship between built-form entropy, income inequality, and local clustering.

\begin{figure}[H] 
    \centering 
    \includegraphics [width=.90\linewidth]{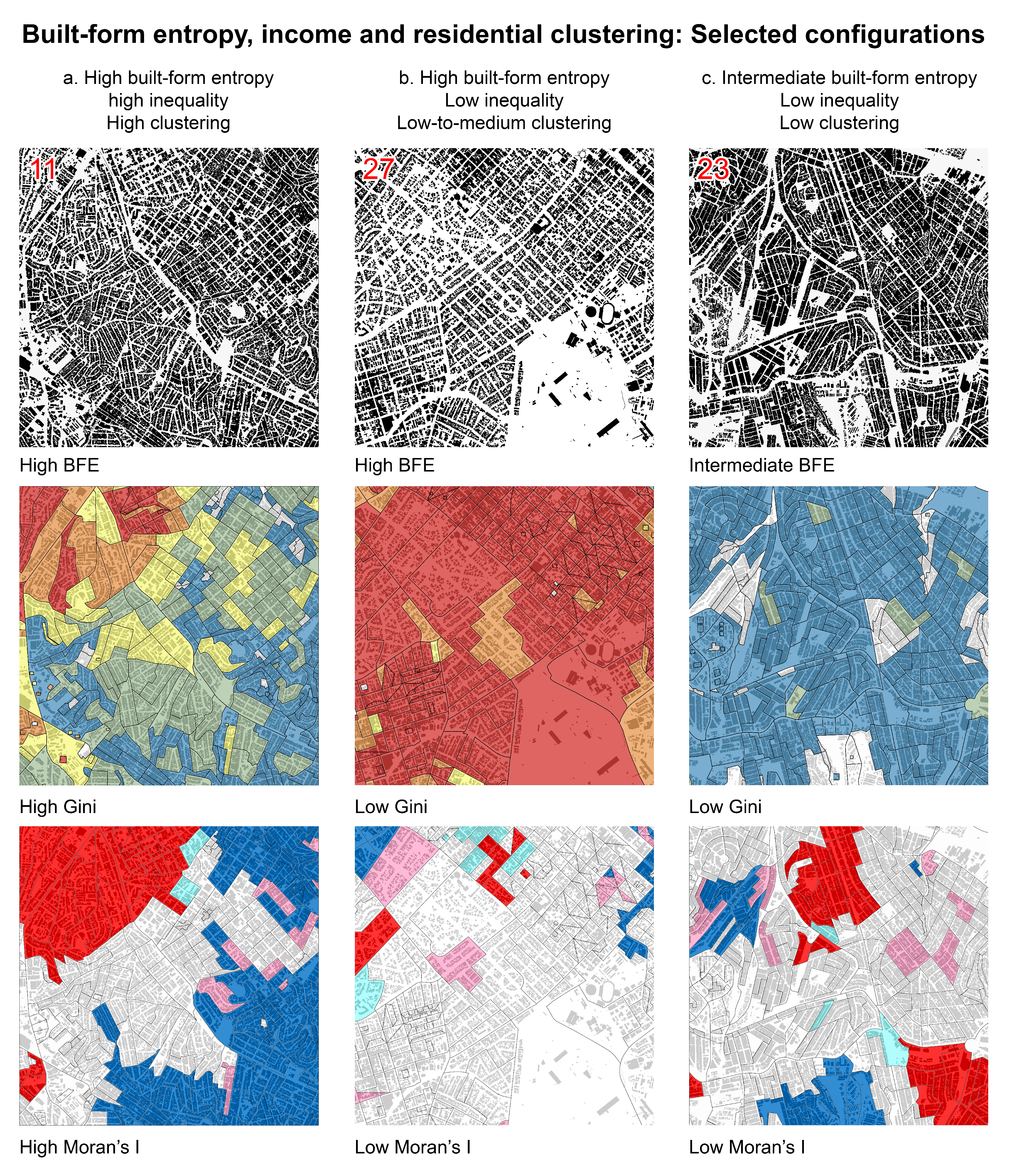} 
    \caption{\small Selected configurations of built-form entropy, income distribution, and residential clustering. Each column shows a 3×3 km cell. (a) Cell 11 illustrates high built-form entropy, high Gini, and strong local clustering, where income clusters co-occur with contrasting morphologies. (b) Cell 27 represents a limiting case of high built-form entropy with a nearly homogeneous high-income group and low internal clustering. (c) Cell 23 shows an intermediate-entropy, lower-income configuration with low inequality and low clustering.}
    \label{fig:Fig_5} 
\end{figure}

\begin{itemize}
    \item[$\bullet$] First, high built-form entropy, high Gini, and high Moran's $I$ correspond most directly to H1: morphologically complex areas in which multiple income groups coexist and cluster locally. We find that very high Moran's $I$ values occur only in high built-form entropy cells, indicating the co-occurrence of distinct income clusters associated with differentiated morphologies at the local scale. 
    In our data, this regime includes cells 11, 12, 14, 17, and 34 (Fig.~\ref{fig:Fig_5}a). Cases where high BFE meets high residential clustering include cells 6, 10, 25 and 62,  at medium Gini values. 
    \item[$\bullet$] Second, a regime of high built-form entropy combined with low Gini and low or low-to-medium Moran's $I$ corresponds most directly to H2. This configuration captures situations where a particular income group concentrates within a specific entropy regime, in an area with relative internal income equality and limited local clustering. It includes lower-income cells 3 and 33, and higher-income cells 19 and 35, and 27 as a limiting case characterised by the dominance of the highest-income group and minimal internal clustering (Fig.~\ref{fig:Fig_5}b).
    \item[$\bullet$] Third, cells with intermediate built-form entropy, low Gini, and low Moran's $I$ correspond to a near-threshold regime 
    in which both inequality and clustering remain comparatively low; 
    this regime includes lower-income cells 
    23, 43, 53, 55, and 56 in eastern S\~ao Paulo (Fig.~\ref{fig:Fig_5}c).

\end{itemize}


A contrasting threshold-related configuration is formed by cells with intermediate built-form entropy, and medium-to-high Moran's $I$ and Gini – namely cells 16, 21, 30, 40, 51, 65, 68, and 71 – showing that areas near the threshold may also coincide with local clustering and uneven income composition (Fig.~\ref{fig:Fig_5}d).
A configuration with low built-form entropy, high Gini and high Moran's $I$ includes only two cells, 47 and 50, showing that ordered built form may also coincide with internal inequality and strong local clustering.
The regimes observed above do not exhaust all possibilities or empirical cases, but represent specific configurations that add nuance and complexity to the broader field defined by the non-linear relationship between built-form entropy and the inequality and clustering measures. 
This trend is also evident at the low-entropy end of this non-linear pattern: configurations with low built-form entropy, medium Moran's $I$, and medium Gini show that ordered built form may also coincide with some internal inequality and local clustering; this regime includes cells 9, 37, 38, 46, 66, and 67, in line with the interpretation that built-form entropy is associated with a structured range of outcomes rather than a single segregation pattern.


\subsection*{Interpreting built-form entropy and residential segregation patterns}

Per-capita income in S\~ao Paulo follows a pronounced centre–periphery gradient radiating from the area southeast of the historic Central Business District (CBD), reflecting a highly segregated urban structure in which income strongly conditions residential location. 
Within this broader pattern, our results indicate systematic associations between built-form entropy and the spatial distribution of income groups (Fig. \ref{fig:Fig_4}c). Income groups do not locate randomly across the built-form entropy (BFE) spectrum; rather, they tend to concentrate in distinct entropy bands: 

\begin{itemize}
    \item[$\bullet$] Low income (blue; 26 cells): Although present across the full BFE range, most cells lie above the threshold $h^T_c\sim 0.317$. Only 1 of 26 (cell 33) occurs above $h_c\sim 0.374$, indicating that low-income locations cluster mainly in such entropy bands (0.317-0.374). 
    \item[$\bullet$] Lower-middle income (green; 20 cells; avg. per-capita income > R\$ 826): 9 of 20 cells lie above $h_c\sim 0.374$, and only 4 of 20 fall below the BFE threshold $h^T_c\sim 0.317$, a tilt toward higher entropy contexts. 
    \item[$\bullet$] Middle income (yellow; 10 cells; avg. per-capita income > R\$ 1436): Spread broadly across the entire BFE range, with no sharp concentration – consistent with a transitional position in both income and morphology. 
    \item[$\bullet$] Upper-middle income (orange; 9 cells; avg. per-capita income > R\$ 2 287): Despite appearing at multiple BFE values, this group concentrates in high-entropy areas: only 3 of 9 upper-income cells fall below $h_c\sim 0.363$, increasing in count as $h_c$ rises. 
    \item[$\bullet$] All highest-income cells (red; 6 cells; > R\$ 3438) lie above $h_c\sim 0.363$ (see Fig. \ref{fig:Fig_4}c).
 \end{itemize}

There is a general association between higher incomes and higher built-form entropy (Fig. \ref{fig:Fig_2}b–c): the probability of observing upper-income cells increases markedly once $h_c$ exceeds 0.363, while low-income cells concentrate above the BFE threshold $h^T_c\sim 0.317$ and are largely absent from the very highest-entropy tail (> 0.374).
The relationship of morphological entropy to income inequality and residential clustering mirrors this pattern, with both Gini and Moran's I following threshold behaviour analogous to that observed for income. \\

\begin{figure}[H] 
    \centering 
    \includegraphics [width=\linewidth]{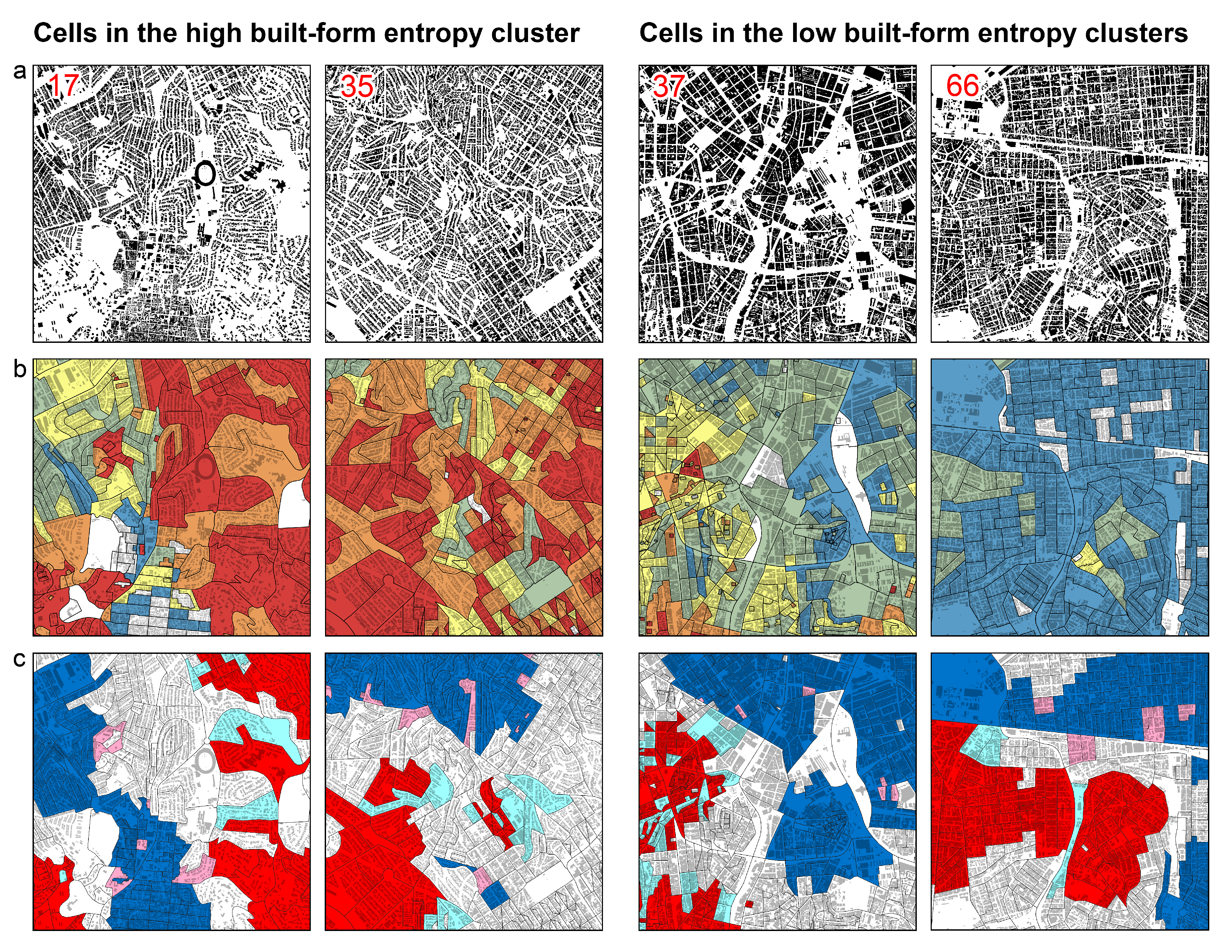} 
    \caption{\small Selected 3×3 km cells from the high-entropy cluster (cells 17 and 35) and the low-entropy clusters (37 and 66). Rows show: (a) built form distributions; (b) tract-level average per-capita income, and (c) local residential clustering of income groups (Moran's $I$). Cell 17 (Vila Sonia - Morumbi) ranks first citywide in BFE, while Morumbi has the highest average per capita income. Cell 35 (Pinheiros–Perdizes) ranks eighth in BFE and eighth in income. Cell 37 (República–Sé) includes the historic CBD; despite a pronounced angular variation in its street network, its compact perimeter blocks with aligned façades yield the lowest BFE in the city. Cell 66 (S\~ao Miguel, Jardim Helena, north of Vila Curuçá) sits on the far eastern periphery – a low-income area with some regular block systems.
    } 
    \label{fig:Fig_6} 
\end{figure}

To ground the broader patterns in concrete urban contexts, Fig.~\ref{fig:Fig_6} contrasts selected cells from the high- and low-entropy clusters. The high-entropy cases include affluent areas such as Morumbi (cell 17), Pinheiros–Perdizes (cell 35), and Jardim Paulista (cell 27), where intricate morphologies coincide either with strong local clustering or with socially homogeneous high-income contexts, as well as mixed or lower-income areas such as Jaguaré (cell 34), where morphological irregularity is also associated with clustered social differentiation. By contrast, the low-entropy cases include the historic CBD in República–Sé (cell 37) and eastern peripheral areas such as São Miguel and Jardim Helena (cells 66–67), where regular block systems and greater morphological continuity coincide with weaker clustering or more moderate internal differentiation. These examples provide empirical illustration for the broader associations described above. More detailed historical and contextual notes on these and other areas are included in the Supplementary Information.

These spatial distributions produce a clear pattern: at the cell scale, built-form entropy is higher where local residential clustering is stronger. Cells with the highest BFE are more likely to show the largest Moran’s $I$ values (Fig.~\ref{fig:Fig_4}d), indicating that high-entropy fabrics tend to coincide with pronounced spatial clustering of income groups within cells.
Income groups tend to occupy distinct entropy ranges. High-income groups are concentrated in high-entropy areas, overwhelmingly located above the estimated entropy threshold at $h^T_c\sim 0.317$. 
These areas display distinct but equally entropic morphologies: low-rise, detached housing embedded in winding street layouts, and vertically intensified fabrics of high-rise buildings. Both forms are typically shaped by planning regimes that privilege parcel-level design over coordinated construction, allowing substantial variation in building footprint and encouraging piecemeal development. In most newly urbanised areas, the requirements for continuity with surrounding street networks are weak, which further contribute to built-form fragmentation. 
Such configurations tend to reduce legibility, permeability, and external accessibility, operating as “soft barriers” that can moderate exposure between social groups (cf. \cite{roberto2021spatial, knaap2024segregated}). 
In contrast, lower-income groups span a range of entropy values but show a clear tendency toward higher entropy, consistent with the morphological variability, including informal urbanisation and self-produced settlements. Most low-income cells lie just above this threshold rather than in the highest-entropy tail, a pattern that becomes more pronounced among lower-middle-income groups. 
Uneven planning histories and incremental urbanisation processes may also funnel low- and lower-middle-income groups into mid-to-high entropy fabrics shaped by partial formalisation, while higher-income groups concentrate in high-entropy environments that offer both distinction and insulation.\\

A notable feature of the threshold-based relationship is the asymmetry across the entropy spectrum. At the high-entropy end, distinct social logics converge on similarly high BFE values: elite enclaving and lower-income settlements each generate morphological irregularity through different urban processes: the former through designed detached housing, high-rise buildings, and winding street layouts, the latter through self-built, incremental, and weakly regulated development. This helps explain why the high-entropy end accommodates both H1-type outcomes, in which multiple income groups coexist and cluster locally, and H2-type outcomes, in which a single income group concentrates in a specific entropy regime and dominates a morphologically complex area.\newline
\indent At the same time, the pattern is not confined to the high-entropy side. At the low-entropy end, both income and residential clustering also rise, albeit less steeply, indicating that relatively affluent areas and moderately clustered configurations may also occur in more ordered morphologies. The asymmetry is twofold: high entropy concentrates a broader range of socio-spatial processes and more intense clustering outcomes, whereas low entropy is compatible with selective concentration and moderate segregation, though through a narrower or weaker set of processes. The same entropy value may therefore index different social arrangements, while similar social outcomes may also arise at opposite ends of the spectrum through different morphological and institutional pathways.\newline
\indent One plausible interpretation for the low-entropy side of this asymmetry is the role of planning in producing ordered environments that are attractive to certain higher-income households, alongside zoning and building rules that contribute to higher land values and selective occupation. Unlike the high-entropy end, where morphological irregularity may operate as a soft barrier, ordered built form may be associated with segregation through regulatory and economic filtering rather than through spatial intricacy itself. These area-specific conditions call for finer-grained contextual research than citywide associational analysis can provide.\\

In turn, the relationship between built-form entropy and inequality also shows non-linearity.
Below the entropy threshold $h^T_c\sim 0.358$, the BFE–Gini relationship is weakly negative, whereas above 
this threshold it becomes positive. Notably, Gini values reach their lowest levels near the threshold, 
suggesting a tipping region in the observed association between morphological entropy and income 
inequality (Fig.~\ref{fig:Fig_4}d; Fig. S2). 
These results suggest that cells containing multiple income groups often coincide with more complex built-form configurations at the local scale. \\

Beyond built-form entropy, our data show that per-capita income, the Gini index, and Moran's $I$ are related in two ways (Fig.~\ref{fig:Fig_4}e). First, both Gini and Moran's $I$ peak at middle income levels: at the extremes, high-income areas tend to exclude low-income residents (and vice versa), which reduces within-cell heterogeneity and local autocorrelation, resulting in lower values for both indices.
Second, identical Gini values can arise from very different spatialisations – dispersed, banded, or tightly clustered – so income composition (Gini) and spatial arrangement (Moran's $I$) need not coincide in principle. However, in the study area the two indices show a moderate positive association (Pearson r $\approx$ 0.59; p < 0.01; Fig.~\ref{fig:Fig_4}f), indicating that income unevenness and local residential clustering tend to co-occur. This relative alignment is consistent with well-established segregation dynamics in which households’ locational preferences – often modest at the individual level – aggregate into spatial clustering of similar groups, as classically illustrated by Schelling’s model of residential segregation \cite{schelling1971dynamic} (see Supplementary Information). Under such conditions, uneven income composition within cells is likely to be spatially expressed through clustering in neighbouring tracts, reinforcing the correspondence between Gini and Moran's $I$. Although this relationship is not logically necessary – since similar levels of inequality can be achieved under different spatial arrangements – it is nonetheless a frequent outcome in settings shaped by income unevenness, land-value gradients, and segregation processes.\\

\section*{Conclusions}

This study set out to clarify how the morphology of the built environment relates to residential 
segregation. By placing built-form entropy (BFE) and residential segregation (RS) metrics on a 
common spatial footing, we find that urban form is systematically associated with income-based 
residential distribution and local clustering, and that these associations are non-linear and threshold-sensitive. The conclusions below synthesise these findings in relation to the complementary hypotheses introduced earlier.  \\
 
(1) Examining the relationship between built-form entropy and residential segregation, we find that highly entropic or irregular morphologies are more frequently observed in areas characterised by enclave-like residential patterns – that is, areas in which income groups co-occur but are internally clustered, with limited mixing across their immediate surroundings and without implying formal closure or institutional enforcement.
Stronger residential clustering is observed in areas with higher built-form entropy (Fig.~\ref{fig:Fig_4}d), in line with our first hypothesis (H1). \\
 
(2) Income groups are not evenly distributed across the built-form entropy spectrum. Instead, they occupy distinct entropy ranges, indicating that the association between built-form entropy and residential segregation varies across income levels (Fig.~\ref{fig:Fig_4}c). 
The clearest pattern is that high-income groups are concentrated in high-entropy areas, overwhelmingly located well above the estimated entropy threshold at $h^T_c\sim 0.317$. In turn, lower-income groups span a broader range of entropy values, but tend to concentrate above the threshold in moderately to highly entropic areas, a pattern even more marked among lower-middle-income groups.
Under certain conditions, these distributions may also produce socially homogeneous areas with low internal clustering. These results are consistent with our second hypothesis (H2).  
\\

(3) 
However, the relationship between built-form entropy and residential segregation, captured by Moran’s $I$, is distinctly non-linear: local clustering reaches its minimum at intermediate entropy values and increases toward both lower and higher-entropy contexts, with a steeper increase at high entropy. 
In substantive terms, the most segregated areas tend to coincide with higher built-form entropy, typically associated with morphologically intricate fabrics composed of fragmented block systems and fine-grained mixtures of building volumes.
The relationships between BFE and 
per-capita income (Fig. 4c) and income inequality (Figs. 4d, S2) show similar non-linear behaviour
(H3).  \\
 

Such patterns unfold within S\~ao Paulo’s pronounced centre–periphery income gradient. On the local scale, within the 3×3 km cells, the relationship between built-form entropy and residential segregation becomes more nuanced. Cells with high entropy frequently display high values of Moran's $I$ – along with the Gini index, indicating that morphological complexity often coincides with strong local clustering and uneven income composition. 
Importantly, both indices reach their lowest values near the entropy threshold, suggesting a tipping region in the morphology–segregation–inequality relationship. Above this point, increasing entropy is associated with greater within-cell inequality and a stronger spatial clustering. \\

The asymmetry of the piecewise regression, with a steeper slope at the high-entropy end than at the low-entropy end, together with the different dispersion of data points at these extremes, suggests that distinct socio-spatial processes converge on related entropy signatures. Elite enclaving may associate high built-form entropy with high-income clustering, while weakly regulated urbanisation may associate similarly high entropy with lower-income clustering.
By contrast, the low-entropy end appears to reflect a narrower set of processes plausibly linked to planning, amenity attraction, and regulatory filtering. 
The high-entropy flank presents a larger dispersion of data points, which may reflect the broader range of socio-spatial processes converging there, and a steeper slope, suggesting that some of these forces may be more intense.
This interpretation helps explain why soft barriers of reduced permeability and intervisibility may be more strongly expressed at the high-entropy end than attractions and planning constraints associated with more ordered built form.\\

These effects align with our general proposition that morphological entropy is associated with spatial conditions under which social segregation is expressed as residential clustering. Under a null expectation of no systematic spatial alignment, the placement of income groups across the built-form entropy spectrum would show no systematic structure, and our metrics would not 
vary coherently with entropy. 
Built form appears to be associated with patterns of social distinction between areas and social homogeneity within them, while potentially providing reduced visibility, accessibility, and intergroup exposure (cf. \cite{hillier1984social, roberto2018spatial} on street networks). 
In this sense, our findings are consistent with an interpretation of urban morphology not as a passive backdrop, but as an active component in the social organisation of residential space – both signal and affordance of segregation. 
In particular, the co-occurrence of high entropy, high average income, and high Moran’s $I$ suggests that built form may help structure residential distributions at this scale rather than merely reflect them.
Together, citywide income gradients and localised morphological variation signal scale-dependent processes: micro-level residential clustering coexists with macro-level income-based residential distribution, adding complexity to the interaction of segregation processes and urban form. \\

At the same time, our results 
focus on the spatial aspect. They document systematic co-location rather than causal effects, and built form may be both a cause and a consequence of segregation dynamics. 
The consistent turning point around $h^T_c\sim 0.317$ highlights threshold sensitivity and calls for alternative analyses of urban form. 
Future work should examine whether this threshold is stable across cities with different built-form structures, income distributions, and planning histories, or whether it is a property specific to S\~ao Paulo's particular morphological and socio-economic configuration.
The analysis nonetheless remains subject to scale and data sensitivities. Results may vary with the modifiable areal unit problem (MAUP). We work with a 3×3 km lattice. Cell sizes were tested and 
the applied cell size shows the required robustness in both entropy measurement and local sampling 
of census tracts. Areal-weight income transfers from census tracts, so the position of the tessellation 
can slightly shift Gini, Moran's $I$ and their correlations with $h_c$. 
The tessellation origin effect – moving the grid a few hundred metres – can change which tracts fall into which cells and thus alter within-cell heterogeneity and local clustering. Notwithstanding, effects of shifts in tessellation position are likely to be balanced out within the overall pattern of income and BF distribution. \\


Finally, the study is associational: piecewise fits and correlations indicate systematic co-location, not causation. However, the consistent co-location of high built-form entropy and strong residential segregation clearly go beyond what a random allocation would produce.
Two non-exclusive mechanisms are plausible for explaining these effects: selection, in which income groups preferentially locate in high-entropy fabrics to limit visibility, access, and contact; and production, in which groups  (and institutions) shape morphology to achieve the same effects. Identifying the dominant pathway lies beyond this study, but the overlap suggests that morphology may be actively involved in segregation dynamics.
Future work should replicate the study across multiple cities and time slices in comparative and longitudinal tests (e.g. pre/post zoning changes) to 
assess transferability and contextual dependence. 
Even so, the observed patterns point to potential practical levers in planning agendas: changes in urban block arrangement and permeability, parcel structure, and building configuration may modify the spatial conditions associated with segregation, even where citywide income gradients persist. 
An applied research path involves examining the mediating role of planning interventions such as zoning reforms and redevelopment policies, addressing questions like: can morphology guided by planning lead to changes in residential segregation – say, may the adoption of more systemic, form-based codes prioritising spatial criteria such as continuity in urban block systems and building placement decrease  morphological entropy? Such efforts could also clarify why similar income mixes produce different segregation outcomes under different physical fabrics, and whether targeted morphological interventions can meaningfully reconfigure clustering, exposure, and segregation dynamics. \\

Beyond the empirical findings, this study aims to contribute in four main ways. First, it brings the physical fabric into residential segregation analysis as a structural property examined alongside inequality and segregation indices. Second, it identifies the form of association between built form and residential segregation, including thresholds that indicate turning points in the relationship. Third, it addresses scale dependence, showing citywide income gradients and local clustering under different morphological entropy regimes. Finally, it offers a replicable workflow, from building footprints to built-form entropy, segregation indices, and association tests, that can be applied to other cities and contexts.\\

\section*{Methods}

We propose a measurement strategy that puts morphology and segregation on the same footing. We 
use a regular tessellation to place all measures on a common, equal-area spatial support. This choice 
serves four purposes. (i) Harmonisation: built-form entropy (from building footprints) and residential 
segregation (from census tracts) are reported on incompatible geometries; the grid provides a 
neutral carrier that avoids privileging administrative boundaries. (ii) Comparability: equal-area cells 
make cross-city and within-city comparisons meaningful (no size/shape biases), reducing one facet of 
the MAUP. (iii) Statistical robustness: Built-form entropy is computed from building footprints and, therefore, requires a tessellation with sufficiently large cells to ensure reliable entropy estimates. In our study, 3×3 km cells provide a balance between spatial resolution and sample size: each cell 
contains enough buildings to estimate the entropy reliably and includes a substantial number of census tracts to support within-cell segregation measures. Across S\~ao Paulo, cells contain between 71 and 
386 census tracts, with an average of 187 tracts per cell, providing adequate sampling for 
estimation of both the Gini index and Moran's $I$. 
Sensitivity checks using alternative cell sizes and grid offsets indicated that smaller cells make entropy estimation impractical 
and clustering estimates fragile, while larger cells dilute local segregation patterns, supporting the choice of the 3×3 km tessellation as a pragmatic compromise. 
(iv) Reproducibility and sensitivity: a regular lattice is simple to replicate and 
to stress-test (offset/jittered grids, alternative cell sizes, or hexagons) to assess tessellation sensitivity. In short, this approach aims at improving comparability and statistical power.\\

Our approach involves three steps. First, we quantify urban morphology by estimating built form entropy (BFE) from building footprints using a previously proposed 
measure based on Shannon's entropy \cite{brigatti_entropy_2021}. Second, we measure residential segregation (RS) from the spatial distribution of per-capita income using Moran's $I$ (local spatial autocorrelation). Third, we assess the relationship 
between BFE and RS by correlating their variations across the study area (Fig. 1). 
Residential segregation measurement is derived from census tracts; we 
overlay the same grid to ensure comparability and 
obtain a stable depiction of local income composition.

\subsection*{Estimating built-form entropy}

\begin{figure}[!htbp]
    \centering 
  \includegraphics [width=\linewidth]{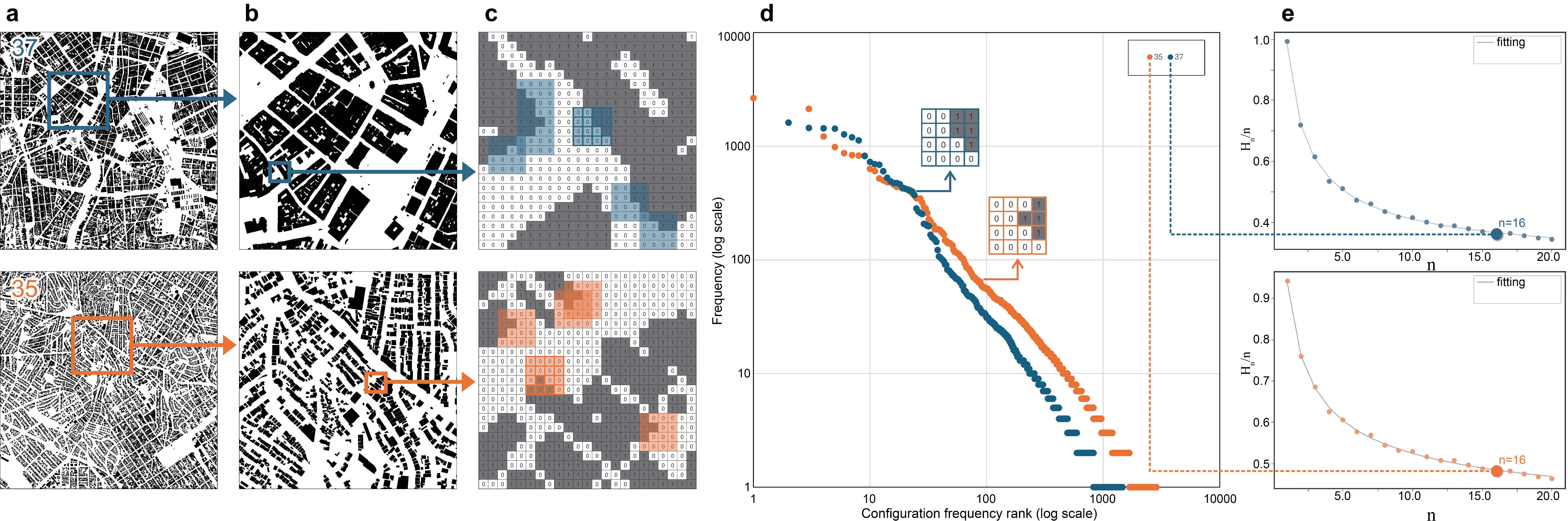} 
    \caption{\small Illustration of the method for estimating built-form entropy (BFE) in cells 37 (top, historic CBD) and 35 (bottom, Pinheiros–Perdizes). (a) The algorithm scans built-form maps in 3×3 km cells and counts the frequencies of local configurations across blocks of different sizes. (b) Zoomed-in views of the building-footprint arrangement. (c) Detection process for a selected configuration with $n$ = 16. Insets outlined in blue (cell 37) and orange (cell 35) highlight examples of size-16 blocks. (d) Rank–frequency distribution of built-form configurations for blocks of size 16. (e) Extrapolation of Shannon entropy to infinite block size. Cell 37 shows more regular arrangements, whereas cell 35 exhibits greater configurational variety.
    } 
    \label{Fig_map} 
\end{figure}

Urban morphology was characterised by reducing the built form to two-dimensional arrangements 
based on building footprints. These figure-ground diagrams, also known as Nolli maps, can be 
explored to analyse and study the urban form based on built/unbuilt distinctions. Building footprints were extracted from the S\~ao Paulo municipality map obtained from GeoSampa and OpenStreetMap services. Considering a trade-off between resolution and data availability \cite{brigatti_entropy_2021, netto_urban_2023}, we chose 
to partition the study area into a Cartesian grid with squares corresponding to geographic areas of 9 
$km^2$ (see Fig. \ref{Fig_map}). The image of each square was resized to produce areas covered by 
$10^3$ cells, each representing a space of $3 \times 3$ meters, and converted to a monochrome 
system. Finally, they were mapped into a matrix of size $1000 \times 1000$ cells with binary 
numerical values, which represent the built/unbuilt distinction. In summary, each square is 
represented by a bidimensional matrix where each element can be a zero, if it stands for an open 
space cell, or a one, for a built form cell. We focused our analysis on squares mapping areas of the 
city with high continuity in built form, avoiding large empty areas or rarefied urbanisation patterns 
generally present in the more peripheral metropolitan areas of the city.  Our method for estimating 
entropy is well-fitted for continuous urban areas. The high continuity and relative homogeneity of built form allow us to use a specific extrapolation technique that will prove useful for estimating the 
entropy of two-dimensional sequences. \\

Shannon's definition of entropy is based on probabilistic concepts \cite{shannon1948mathematical}. First, it describes properties of the ensemble of sequences produced by a source. In addition, the source that emits the sequences is characterised by a probability distribution which, in turn, is defined
by 
the set of all possible produced sequences. Since these quantities, in general, are not directly accessible, Shannon already proposed  some methods to overcome this difficulty \cite{shannon1951prediction}. These approaches are based on the idea that the entropy can be numerically estimated by studying only one very long 
sequence. For a one-dimensional sequence (a string), a possible approach relies on defining the block entropy of order $n$ as: 
 
\begin{equation} 
H_n=-\sum_k  p_n(k) \log_{2}[p_n(k)], 
\label{1entropy} 
\end{equation} 
 
where blocks are string segments of size $n$, and the sum runs over all the $k$ possible $n$-blocks. 
Equation~(\ref{1entropy}) corresponds to the Shannon entropy of the probability distribution $p_n(k)$. The Shannon entropy density of the considered system, which we indicate with $h$, is 
obtained from the following limit \cite{schurmann1996entropy}: 
 
\begin{equation} 
h=\lim_{n \to \infty} H_n/n, 
\label{hentropy} 
\end{equation} 
 
which measures the average amount of randomness per symbol that persists after all correlations 
and constraints are taken into account. More details about this method can be found in 
\cite{schurmann1996entropy,cover1991elements}. This approach can be generalised to our 
situation, where we analyse sequences of symbols in two dimensions, as reported in 
\cite{brigatti_entropy_2021,brigatti2022comment}. The decisive point consists of defining the $n$-blocks for a two-dimensional matrix. The most intuitive idea is to consider a block of size $n$ as a square 
containing $n$ cells. The sequences of $H_{n}$ for $n$ values that do not correspond to squares are 
then obtained by considering blocks that interpolate perfect squares. Note that there is no unique natural way to scan a bidimensional matrix. We tested our approach for different reasonable forms of constructing the blocks, and even for randomly constructed blocks. For matrices representing building footprints, the use of different approaches does not significantly influence the estimation of $H_n$  \cite{brigatti_entropy_2021,netto_urban_2023}. In this analysis, we adopt the block definition introduced in \cite{netto_urban_2023}, considering blocks up to $n=20$.
The limit in equation \ref{hentropy} can be empirically obtained by 
fitting the $H_n/n$ points with an appropriate function and then taking its limit for $n \to \infty$. 
We find heuristically that the following ansatz provides an excellent fit \cite{brigatti_entropy_2021}: 
 
\begin{equation} 
H_n/n \approx a+c/n^b, \qquad  b,c>0. 
\label{fitting} 
\end{equation}

The fitted value of $a$ gives a reasonable extrapolation of the Shannon entropy $h$. Specific considerations on the interpretation and application of this method for characterising the entropy of urban fabric can be found in \cite{netto2018cities,brigatti_entropy_2021,netto_urban_2023}, generalisations and technical details 
in \cite{desouza2022entropy,brigatti2022comment}. By applying this approach, we obtain the value of the Shannon entropy of each square that tessellates the city of S\~ao Paulo. Our city contains squares with varying 
densities of built-form cells. If the density value is far from 50\%, there is an important reduction in 
the entropy value caused just by this asymmetry. This fact is not necessarily connected with randomness and correlations in the built form. For example, the presence of a river or an urban park, which are uniform unbuilt areas, can sensibly reduce the entropy value. As we are interested in 
characterising the randomness of built form, we introduce a heuristic approach for tackling this problem: we correct the entropy value $h$ by adding the term that corresponds to the reduction of 
entropy due to the asymmetry in the frequency of unoccupied (0) and occupied (1) cells present in the data. Since $H_1$ is the entropy of the single-cell distribution, the reduction in the value of $h$ due to this contribution is equal to $1-H_1$. Adding this term to $h$ means that, if we do not consider 
the asymmetry in 0 and 1, the system appears to be more random than it is by an amount $1-H_1$. Such a term will be null in a matrix with perfect equiprobability between 0 and 1. It follows that our corrected entropy is \cite{brigatti_entropy_2021}: 
 
\begin{equation} 
h_c= h+(1-H_1). 
\end{equation} 
 
This procedure does not correct all the contributions that an asymmetry in the frequency of 0 and 1 makes to the value of $h$ but, at least, it takes away the larger ones. 
The way $h$ differs from $H_1$ captures in an integrated and involved form how the presence of correlations determines the 
effective randomness \cite{brigatti_entropy_2021}.\\ 

Finally, we applied a Hot Spot Analysis (Getis–Ord \textit{Gi}*) \cite{getis1992analysis} of built-form entropy (BFE) values computed for 3×3 km grid cells to identify statistically significant clusters of high entropy (hot spots) and low entropy (cold spots) (Fig. 2b). 
The Getis–Ord statistic compares the sum of BFE values of each cell and its neighbouring  cells to the global distribution of BFE values. Cells with significantly higher-than-expected local sums 
are identified as hot spots (red), while those with significantly lower-than-expected sums are identified as cold spots (blue). Statistical significance is evaluated using z-scores (with multiple-testing correction), indicating spatial concentrations of high or low entropy beyond what would be expected under a null model of spatial randomness. 

\subsection*{Estimating residential segregation}

Social groups were characterised using census-tract population and per-capita income for the municipality of S\~ao Paulo from the 2010 Brazilian Census (IBGE) \cite{IBGE2010Censo}. We used the average per-capita 
income reported for each tract as our socioeconomic indicator. Figure 2s in the Supplementary Information
illustrates how census tracts relate to a single 3×3 km grid cell in our analysis. We also calculated the Gini index for each grid cell. The Gini coefficient is the most commonly used inequality index. It is a measure of statistical dispersion that, in our case, represents the degree of inequality of income distribution. It is used to characterise how far an income distribution deviates from an equal distribution and ranges from 0 to 1, with 0 indicating perfect equality and 1 representing perfect inequality. We computed Gini coefficients with the R {\it DescTools} package \cite{DescTools} using the tract-level per-capita income values  for all tracts intersecting each grid cell. \\
 
 
Income serves as a proxy for social grouping, so looking at an income map may suggest where segregation is higher or lower. To quantify this intuition, we explored a spatial association index classified by Reardon and O'Sullivan \cite{reardon_measures_2004} in the ‘spatial evenness’ category. Moran's $I$ is a spatial autocorrelation measure that assesses whether an attribute’s values show clustering, randomness, or dispersion across neighbouring areas. It is well suited to analyse residential segregation, typically characterised by spatial clustering of similar groups (Fig. S3b in the Supplementary Information). In the context of residential segregation research, Moran's was first used by Frank \cite{frank2003moran} to characterise socioeconomic and racial residential patterns in U.S. cities. 
Using the function {\it moran.test} of the {\it Spdep} R package \cite{spdep2018, spdep2024}, we computed Moran's $I$ for each census tract, providing a measure of the local Moran statistic. 
A global Moran's coefficient is obtained for each square by averaging all these local values. 
We first specified the neighbouring polygons and spatial weights. Our primary specification uses a fixed-distance band of 500 m, assigning (equal) weights of 1 to all tracts whose centroids fall within that radius. As a robustness check, we also estimated models with contiguity weights (Queen criterion: polygons sharing at least one vertex).  
For comparison, we also derived Local Indicators of Spatial Association (LISA) with local Moran and summarised them within cells, which produced patterns consistent with the global statistics. \\
 
This coefficient can range between -1 and +1. Positive values indicate that as the income of a given polygon increases, the incomes of its neighbouring polygons tend to increase as well, suggesting the presence of a spatial cluster of high or low incomes. Values close to zero imply a random spatial income distribution, with no apparent clustering. Negative values denote a dispersed spatial pattern, where high-income polygons are surrounded by low-income polygons, and vice-versa, indicating a negative spatial autocorrelation. 
The values of the Moran's $I$ index and the census tract 
income estimates can be carried over on the squares of the 3km $\times$ 3km tessellation using a technique for estimating feature values in overlapping but incongruent polygons (Fig. S3c in the SI). We use the 
areal weighted interpolation method implemented in R \cite{interpolation}. This method was also 
used to transfer census tract income estimates to the 9 $km^2$ squares. Such an approach assumes 
the variable is uniformly distributed across each spatial unit (3x3 km grid cell). Because cell boundaries are artificial, results are susceptible to MAUP.

\section*{Acknowledgements}
\vspace{-1em}
\small
The authors disclosed receipt of the following financial support for the research, authorship, and/or publication of this article: Research Centre for Territory, Transport and Environment, University of Porto (CITTA FEUP); Foundation for Science and Technology (Fundação para a Ciência e Tecnologia | FCT, Portugal, grant 2023.07510.CEECIND); Foundation for Research Support (Fundação de Amparo à Pesquisa do Rio de Janeiro | FAPERJ, grant E-26/211.381/2021); and National Council for Scientific and Technological Development (Conselho Nacional de Desenvolvimento Científico e Tecnológico | CNPq, Brazil, grant numbers 315086/2020-3 and 305008/2021-8).

\section*{Supplementary information}
\vspace{-1em}
\small
Supplementary information for “Is segregation encoded in urban form? An entropy-based analysis” includes an extended literature review, grid-level measurement of built-form entropy, income and residential segregation indicators, full piecewise regression results for the built-form entropy–Gini relationship in S\~ao Paulo, and contextual notes on high- and low-entropy clusters:
\url{https://zenodo.org/uploads/19651255}

\section*{Data availability}
\vspace{-1em}
\small
The data supporting the findings of this study are publicly available. Census-tract income data were obtained from the Brazilian Institute of Geography and Statistics (IBGE). Building-footprint data were obtained from OpenStreetMap (OSM) and GeoSampa (S\~ao Paulo City Hall / Municipal Government of S\~ao Paulo). 

\section*{Code availability}
\vspace{-1em}
\small
The code developed for this work will be shared in a public repository following peer review, in line with journal policy.

\section*{References}
\vspace{-4em}
\small
\setstretch{1.0}
\setlength{\bibsep}{0pt}
\setlength{\itemsep}{0pt}
\setlength{\parskip}{0pt}
\bibliographystyle{naturemag}
\bibliography{BFE_vs_RS_Refs_Overleaf_260326}

\end{document}